\documentclass[conference]{IEEEtran}
\usepackage{amsmath,amssymb,amsfonts}
\usepackage{mathtools}
\usepackage{algorithmic}
\usepackage{listings}
\usepackage{graphicx}
\usepackage{textcomp}
\usepackage{xcolor}
\usepackage{xspace}
\usepackage[style=ieee,backend=bibtex,minnames=2,maxnames=3]{biblatex}

\usepackage{ifxetex}
\ifxetex
  \usepackage{fontspec} \else
\fi

\usepackage{microtype}
\usepackage{comment}
\usepackage{balance}
\usepackage{flushend}

\usepackage[font=small,labelfont=bf]{caption}

\usepackage{pgfplots}
\usepackage{pgfplotstable}
\usetikzlibrary{pgfplots.groupplots}
\usetikzlibrary{arrows.meta}
\usetikzlibrary{positioning}
\usetikzlibrary{patterns}

\newcommand\TS{\rule{0pt}{2.6ex}}        
   \newcommand{\hl}[1]{\textbf{{#1}}} \newcommand{\sect}[1]{Section~\ref{#1}\xspace}  \newcommand{\tbl}[1]{TABLE~\ref{#1}\xspace}  \newcommand{\fig}[1]{Fig.~\ref{#1}\xspace}   \newcommand{\eq}[1]{(\ref{#1})\xspace}   
\newcommand{\arbor}[0]{Arbor\xspace}
\newcommand{\nrn}[0]{NEURON\xspace}
\newcommand{\nmodl}[0]{NMODL\xspace}

\newcommand{\daintmc}[0]{Daint-mc\xspace}
\newcommand{\daintgpu}[0]{Daint-gpu\xspace}
\newcommand{\tave}[0]{Tave-knl\xspace}
\newcommand{\corenrn}[0]{CoreNEURON\xspace}
\newcommand{\hbp}[0]{HBP\xspace}

\begin{document}

\title{
    \arbor{} -- a morphologically-detailed neural network simulation library for contemporary high-performance computing architectures
}

\iftrue
\author{
    \IEEEauthorblockN{Nora Abi Akar}
    \IEEEauthorblockA{\textit{Scientific Software \& Libraries} \\
    \textit{CSCS}\\
    Z\"urich, Switzerland \\
    nora.abiakar@cscs.ch}\\
    \IEEEauthorblockN{Anne K\"usters}
    \IEEEauthorblockA{\textit{SimLab Neuroscience, JSC, IAS} \\
    \textit{Forschungszentrum Jülich}\\
    J\"ulich, Germany \\
    a.kuesters@fz-juelich.de}
\and
    \IEEEauthorblockN{Ben Cumming}
    \IEEEauthorblockA{\textit{Scientific Software \& Libraries} \\
    \textit{CSCS}\\
    Z\"urich, Switzerland \\
    bcumming@cscs.ch}\\
    \IEEEauthorblockN{Wouter Klijn}
    \IEEEauthorblockA{\textit{SimLab Neuroscience, JSC, IAS} \\
    \textit{Forschungszentrum Jülich}\\
    J\"ulich, Germany \\
    w.klijn@fz-juelich.de}\\    
    \IEEEauthorblockN{Stuart Yates}
    \IEEEauthorblockA{\textit{Scientific Software \& Libraries} \\
    \textit{CSCS}\\
    Z\"urich, Switzerland \\
    yates@cscs.ch}
\and
    \IEEEauthorblockN{Vasileios Karakasis}
    \IEEEauthorblockA{\textit{Scientific Computing Support} \\
    \textit{CSCS}\\
    Lugano, Switzerland \\
    karakasis@cscs.ch}\\
    \IEEEauthorblockN{Alexander Peyser}
    \IEEEauthorblockA{\textit{SimLab Neuroscience, JSC, IAS} \\
    \textit{Forschungszentrum Jülich}\\
    J\"ulich, Germany \\
    a.peyser@fz-juelich.de}
}

 \fi

\maketitle

\begin{abstract}
    We introduce \arbor, a performance portable library for simulation of large networks of multi-compartment neurons on HPC systems.
\arbor is open source software, developed under the auspices of the \hbp.
The performance portability is by virtue of back-end specific optimizations for x86 multicore, Intel KNL, and NVIDIA GPUs.
When coupled with low memory overheads, these optimizations make \arbor an order of magnitude faster than the most widely-used comparable simulation software.
The single-node performance can be scaled out to run very large models at extreme scale with efficient weak scaling.
 \end{abstract}


\section{Introduction}

\includecomment{intro}
\begin{intro}
\noindent
{\footnotesize ``What I cannot create, I do not understand.''\\[-0.2\baselineskip]\makebox[\linewidth][r]{-- Feynmann's blackboard, February 1988}}
\vskip0.5\baselineskip

The electrical basis of neuronal activity was experimentally
established in 1902 by Bernstein \cite[][Review]{pallotta-1992} with preliminary
chemical models of ionic electrodiffusion. Further refinement by
Hodgkin, Huxley, Katz and Stämpfli of the connection between ionic
flow, electrical activity, and axonal firing quickly followed in the
wet lab with the classical series of experiments on squid giant
axons. However, the crucial breakthrough in understanding the link
between biological activity and electrical activity in neuronal
networks was the theoretical description by Hodgkin and Huxley in 1952
in mathematical and engineering language which allowed the computation
of the depolarization and repolarization of neurons from a few
experimentally accessible parameters. Full multicompartment models
handling complex, bifurcating dendritic arbors
\cite{rall-1962} and assuming binary axonal firing and
velocity allowed for the mathematical construction of entire
neurons. With the addition of synaptic models in the ensuing decades, the fundamental formal description of biological
neuronal networks was captured at the scale of cellular behavior.

The evolution of computing equipment ranging from the desktop PC to
supercomputing centers has enabled a plethora of tools for numerically
computing predictions of neuronal network behavior that is comparable
with a variety of experimental results, thus allowing the rigorous
testing of possible functional models with varying levels of
experimental verification, mathematical validity and stability, and
computational performance.  In 1984, the publication of the Thomas
algorithm for neurons and its implementation in CABLE and \nrn were
the computational breakthrough for efficient computations of complex
dendritic arbors and the diffusion of computational simulation to the
wider neuroscientific community \cite{Carnevale2006}. Continuing
improvements to \nrn and the addition of automated code generation for
neuron models to allow the injection of efficient C code by
neuroscientists not versed in programming made \nrn the premier
neuronal network simulator for multicompartment neurons.

However, the field of neuronal network simulators is diverse,
encompassing point neuron simulators like NEST, interpreted simulators
for fast prototyping like BRIAN, and high-resolution simulators like
GENESIS \cite[][Review]{brette-2007}. As the field expands its ambitions from
small neuronal networks towards the $10^{11}$ neurons in a human
brain, and beyond by a further two orders of magnitude to include
electrically active glial cells, issues of parallel computational
performance have begun to dominate the field \cite{brette-2007},
featuring experimental simulators like SPLIT. Both \nrn and NEST have
been ported to MPI for large clusters and high-performance computing
(HPC) centers; petascale-capable code has been developed as \corenrn,
NEST 4G and \arbor. Use cases have included the neocortical
simulations \cite{markram-2015}, macaque visual cortical areas
\cite{schmidt-2018}, and olfactory bulb simulations
\cite{migliore-2015}.

New HPC architectures such as the addition of ubiquitous GPU resources
have been a new challenge, requiring new code adaptation with codes
such as \corenrn and GeNN for single-node GPU neuronal
networks. Developing performant algorithms for computing the Hines
matrix on GPUs and other vectorize hardware has been an additional
hurdle \cite{VALEROLARA2017566,huber-2018}. The development of \arbor \cite{akar-2018} has focused on
tackling issues of vectorization and emerging hardware architectures
by using modern C++ and automated code generation, within an
open-source and open-development model.

\end{intro}

\section{Model}
\label{sec:model}
\arbor simulates networks of spiking neurons, and in particular,
networks of multi-compartment neurons.

In this model, the only interaction between cells is mediated
by spikes\footnote{Interaction via gap junctions is under
development in \arbor, but is not currently supported.},
discrete events initiated by one neuron which are
then propagated and distributed to typically many other neurons
after a delay. This abstracts the process by which an action
potential is generated by a neuron and then carried via an
axon to a number of remote synapses.

For simulation purposes, this interaction between neurons is
modeled as a tuple: source neuron, destination neuron,
synapse on the destination neuron, propagation time, and a
weight. The strictly positive propagation times allow for
semi-independent evolution of neuron state: if $\Delta T$
is the minimum propagation time, then the integration of
the neuron state from time $t$ to $t+\Delta T/2$ only needs
spikes that were generated at $t<t-\Delta T/2$.

While \arbor supports a number of neuron models, the main focus is
on the simulation of multi-compartment neurons. Here each
cell is modeled as a branching, one dimensional electrical
system with dynamics governed by the cable equation, with ion
channels and synapses represented by additional current sources.
(See \cite{HodgkinRushton1946} for an early example of this
model applied to a crustacean axon.)
The domain is discretized (hence the term multi-compartment)
and the consequent system of differential equations is integrated
numerically. Spikes are generated when the membrane voltage
at a given point --- typically on the soma --- exceeds a fixed
threshold.

\begin{figure}[t]
    \centering
    \begin{tikzpicture}[scale=0.9]
	\foreach \r in {-1,1}
	    \foreach \y in {0,-0.1cm}
		\draw[yscale=\r] plot[smooth,yshift=\y] coordinates{(0,1) (1,1.3) (4,1.1) (6,1.5)};
	\draw (0,0) ellipse [x radius=0.2cm, y radius=1cm];
	\draw (0,0) ellipse [x radius=0.18cm, y radius=0.9cm];
	\draw (6,-1.5) arc [start angle=-90, end angle=90, x radius=0.3cm, y radius=1.5cm];
	\draw (6,-1.4) arc [start angle=-90, end angle=90, x radius=0.28cm, y radius=1.4cm];
	\draw[dotted] (6,1.5) arc [start angle=90, end angle=270, x radius=0.28cm, y radius=1.4cm];
	\draw[-Latex] (0.5,0) -- node[above, near end] {\tiny $J_1$} (-1,0);
	\draw[-Latex] (7,0) -- node[above, near start] {\tiny $J_2$} (5.5,0);
	\draw[-Latex] (3,-0.5) -- node[left, near end] {\tiny $J_\text{membrane}$} (3,-2);
    \end{tikzpicture}
    \caption{Current balance in dendrite section, $J_\text{membrane}=J_2-J_1$.
	The membrane is treated as a leaky capacitor
	with extra current sources, surrounding an ohmic conductor of variable radius.}
    \label{fig:currentbalance}
\end{figure}
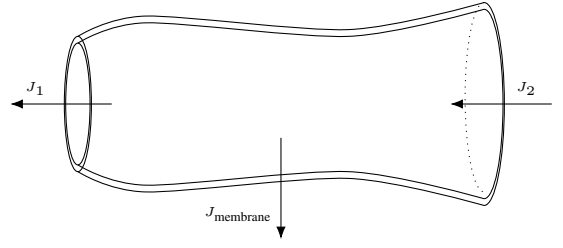

The cable equation is derived from the balance of trans-membrane
currents with the axial currents that travel through the
intracellular medium (\fig{fig:currentbalance}). Membrane currents
comprise a mixture of discrete currents and distributed current
densities, derived from: the membrane capacitance; a surface
density of ion channels; and point sources from synapses and
experimentally-injected currents. The synapse and ion channel
currents themselves are functions of local states that
are in turn governed by a system of voltage-dependent ODEs.

The resulting equations have the form
\begin{subequations}
    \begin{align}
	\begin{split}
	    \frac{\partial}{\partial x}\Bigl(\sigma\frac{\partial v}{\partial x}\Bigr)
	    &=
	    \Bigl(c_m\cdot\frac{\partial v}{\partial t} +
		\smashoperator{\sum_{\text{channels $k$}}}
		    g_k(\underline{s}_k(x, t))(v-e_k^\text{rev})\Bigr)\cdot
	    \frac{\partial S}{\partial x}
	    \\
	    &\quad+
	    \smashoperator{\sum_{\text{synapses k}}}
		I_i^\text{syn}(\underline{s}_k^\text{syn}(t), v(x_k^\text{syn}))\,\delta x_k^\text{syn}
	    \\
	    &\quad+
	    \smashoperator{\sum_{\text{injections k}}}I_k^\text{inj}(t)\,\delta x_k^\text{inj},
	\end{split}
	\\
	\frac{d}{dt}\underline{s}_k(x, t)
	&=
	f_i(\underline{s}_k, v(x, t)),
	\\
	\frac{d}{dt}\underline{s}_k^\text{syn}(t)
	&=
	f_i^\text{syn}(\underline{s}_k^\text{syn}, v(x_k^\text{syn}, t), t),
    \end{align}
\end{subequations}
where $\sigma$ is the axial conductivity of the intracellular medium;
$c_m$ is the membrane areal capacitance; $g_k$ is the
areal conductance for an ion channel of type $k$ as a function
of an ion channel state $\underline{s}_k$, with
corresponding reversal potential $e_k^\text{rev}$;
$S(x)$ is the membrane surface area as a function of axial distance $x$;
$I_k^\text{syn}$ is the current produced by a synapse at position $x_k^\text{syn}$
as a function of the synaptic state $\underline{s}_k^\text{syn}$ and local voltage;
and $I_k^\text{inj}(t)$ is the injected current at position $x_k^\text{\rlap{inj}}$.
The functions $f_k$ and $f_k^\text{syn}$ encode the biophysical properties of
the ion channels and synapses, and will vary from simulation to simulation.

\arbor performs a vertex-centered 1-D finite volume discretization in space, using a
first-order approximation for the axial current flux. This provides a coupled set
of ODEs for the discretized voltages $V_i$ for each control volume $X_i$, the
synapse states $\underline{s}_k^\text{syn}$, and the discretized states of the
ion channels $\underline{s}_{k,i}$ in $X_i$:
\begin{subequations}
    \begin{align}
	\begin{split}
	    \label{eq:discrete-v}
	    c_i\frac{dV_i}{dt}
	    &=
	    \smashoperator{\sum_{j:\ X_j\cap X_i\neq\emptyset}} \sigma_{i,j}(V_j-V_i)
	    -
	    \smashoperator{\sum_{k:\ x_k^\text{inj}\in X_i}} I_k^\text{inj}(t)
	    \\
	    &\quad-
	    \smashoperator{\sum_{k:\ x_k^\text{syn}\in X_i}}
		I_k^\text{syn}(\underline{s}_k^\text{syn}, V_i)
	    \\
	    &\quad-
	    \smashoperator{\sum_{\text{channels $k$}}}
		S_i\cdot g_k(\underline{s}_{k,i})(V_i-e_k^\text{rev}),
	\end{split}
	\\
	\frac{d\underline{s}_{k,i}}{dt}
	&=
	f_k(\underline{s}_{k,i}, V_i),
	\\
	\frac{d\underline{s}_k^\text{syn}}{dt}
	&=
	f_k^\text{syn}(\underline{s}_k^\text{syn}, V_i, t),
    \end{align}
\end{subequations}
where $c_i$ is the integrated surface capacitance on $X_i$, $S_i$ is the surface area of $X_i$,
and $\sigma_{i,j}$ is the reciprocal of the integrated axial resistance between the
centers of $X_i$ and $X_j$.

The ODEs are solved numerically by splitting the integration of the voltages $V_i$ from
the integration of the ion channel and synapse states $\underline{s}_{k,i}$,
$\underline{s}_k^\text{syn}$: at a time $t$, the synapse, channel and injection currents
are calculated as a function of $V_i(t)$; $V_i(t+\delta t)$ is then computed from
\eq{eq:discrete-v} by the implicit Euler method; finally, the state variables are integrated
with the updated voltage values. The method of state integration will depend upon the
particular set of ODEs used to describe their dynamics.

Synapse state is at best only piece-wise continuous: the arrival of
a spike at the synapse can cause a step change in its state, and correspondingly in
its current contribution. The time step $\delta t$ used for integration is shortened
as necessary to ensure that these updates can all be accounted for at the beginning
of an integration step.

For any given $\delta t$, the implicit Euler step involves the solution of the linear
system
\[
    \frac{c_i}{\delta t} V_i' + \sum_j \sigma_{i,j} V_i' - \sum_j \sigma_{i,j} V_j'
    = -I_i^\text{memb} + \frac{c_i}{\delta t} V_i,
\]
where $\sigma_{i,j}$ is non-zero only for neighboring control volumes $X_i$ and $X_j$.
The corresponding matrix on the left hand side is symmetric positive definite, comprising
a weighted discrete Laplacian on the adjacency graph of the control volumes plus a
strictly positive diagonal component given by $c_i/\delta t$.

By appropriate numbering of the control volumes, this matrix becomes nearly tridiagonal,
with other non-zero terms corresponding to branching points in the dendritic tree,
and can be solved efficiently via an extension of the Thomas algorithm
(see \cite{Hines1984}; matrices of this form are now often termed Hines matrices.)

\section{Design}
\label{sec:design}
\arbor is designed to accommodate three primary goals: scalability; extensibility;
and performance portability.

Scalability is achieved through distributed model construction, following the
abstraction of a \emph{recipe} as described below, and through the use of an
asynchronous MPI-based spike communication scheme. \arbor is extensible in that
it allows for the creation of new kinds of cells and new kinds of cell
implementations, while target-specific
vectorization, code-generation and cell group implementations allow
hardware-optimized performance of models specified in a portable and generic
way.

\subsection{Cells and recipes}
\label{sec:design-groups}

The basic unit of abstraction in an \arbor model is the \emph{cell}. A cell represents
the smallest model that can be simulated, and forms the smallest unit of work
that can be distributed across processes. Cells interact with each other only
via spike exchange.

Cells can be of various types, admitting different representations and
implementations. \arbor currently supports specialized leaky integrate and fire
cells and cells representing artificial spike sources in addition to the
multi-compartment neurons described in \sect{sec:model}.

Large models, intended to be run on possibly thousands of nodes, must be able
to be instantiated in a distributed manner; if any one node has to enumerate
each cell in the model, this component of the simulation is effectively
serialized. \arbor uses a \emph{recipe} to describe models in a cell-oriented
manner. Recipes derive from a base recipe class and supply methods to map a
global cell identifier to a cell type, a cell description, and to a list of all
connections from other cells that terminate on it. This provides a deferred
evaluation of the model, allowing the process of instantiation of cell data to
be performed in parallel both on and across ranks.

\subsection{Cell groups}

A \emph{cell group} is an object within the simulation that represents a collection of
cells of the same type together with an implementation of their simulation. As
an example, there are two cell group implementations for multi-compartment
cells: one for CPU execution, which runs one or a small number of cells in
order to take advantage of cache-locality; and a CUDA-based
implementation for GPUs which runs a very large number of cells in order to utilize
the very wide parallelism of the device.

The partitioning of cells into cell groups is provided by a \emph{decomposition}
object; the default partitioner will distribute cells of the same type evenly
across nodes, utilizing GPU-based cell groups if they are available. Users,
however, can provide more specialized partitions for their models.

The \emph{simulation} object itself is responsible for managing the instantiation of
the model and the scheduling of the spike exchange task and the integration
tasks for each cell group.

\subsection{Mechanisms}

The description of multi-compartment cells also includes the specification of
ion channel and synapse dynamics. In the recipe, these specifications are
called mechanisms and are referenced by name; the cell group will select a
CPU- or GPU-implementation for the mechanism as appropriate from a default or
user-supplied mechanism catalog.

Mechanisms in the cataloger can be hand-coded for CPU or GPU execution, but are
more typically compiled from a high-level domain specific language. \arbor
provides a transpiler called \texttt{modcc} that compiles a subset of the \nrn mechanism
specification language \nmodl to architecture-optimized vectorized C++
or CUDA source. These can then be compiled and linked in to an \arbor{}-using
application for use in multi-compartment simulations.

\section{Implementation}
The \arbor library is written in C++14 and CUDA. The implementation has been
driven by a philosophy of `simple by default': the code makes extensive use of
the C++ standard library data structures, and more complicated solutions are
used only where optimization is necessary. External dependencies are minimized:
\arbor includes its own thread pool implementation and vectorization library,
and in general a third-party library will only be employed if the burden of
writing or maintaining a native solution becomes infeasibly large.

We adopt an open development model, with code, bug reports and issues hosted on
GitHub \cite{akar-2018}.

\subsection{Simulation workflow}
\label{sec:impl-workflow}

\begin{figure*}[htp!]
    \begin{tikzpicture} [scale=0.7, every node/.style={scale=0.7}]
        \tikzstyle{thread}=[draw=none, rectangle, fill=white, minimum height=0.5cm, minimum width=0.1cm, anchor=west]
\tikzstyle{cell}=[thick, draw=brown, rectangle, minimum height=0.5cm, fill=brown!5, anchor=north west,pattern=crosshatch dots,minimum width=0.8cm]
\tikzstyle{enqueue}=[thick, draw=olive, rectangle, minimum height=0.5cm, fill=olive!5, anchor=north west,minimum width=0.3cm]
\tikzstyle{exchange}=[thick, draw=red, rectangle, minimum height=0.5cm, fill=red!5, anchor=north west,pattern=vertical lines,minimum width=0cm]
\tikzstyle{walk}=[thick, draw=blue, rectangle, minimum height=0.5cm, fill=blue!5, anchor=north west,pattern=north east lines,minimum width=0cm]

\node[exchange, minimum width=0.85cm](tsk_0_0) at (0.000000cm,-0.000000cm){};
\node[walk, minimum width=2.15cm](tsk_0_1) at (0.900000cm,-0.000000cm){};
\node[cell, minimum width=0.75cm](tsk_0_2) at (3.100000cm,-0.000000cm){};
\node[enqueue, minimum width=0.25cm](tsk_0_3) at (3.900000cm,-0.000000cm){};
\node[enqueue, minimum width=0.25cm](tsk_0_4) at (4.200000cm,-0.000000cm){};
\node[enqueue, minimum width=0.25cm](tsk_0_5) at (4.500000cm,-0.000000cm){};
\node[exchange, minimum width=0.95cm](tsk_0_6) at (5.050001cm,-0.000000cm){};
\node[walk, minimum width=1.75cm](tsk_0_7) at (6.050001cm,-0.000000cm){};
\node[cell, minimum width=0.75cm](tsk_0_8) at (7.850000cm,-0.000000cm){};
\node[enqueue, minimum width=0.25cm](tsk_0_9) at (8.650001cm,-0.000000cm){};
\node[enqueue, minimum width=0.25cm](tsk_0_10) at (8.950001cm,-0.000000cm){};
\node[enqueue, minimum width=0.25cm](tsk_0_11) at (9.250001cm,-0.000000cm){};
\node[enqueue, minimum width=0.25cm](tsk_0_12) at (9.550001cm,-0.000000cm){};
\node[exchange, minimum width=2.05cm](tsk_0_13) at (10.000002cm,-0.000000cm){};
\node[walk, minimum width=4.25cm](tsk_0_14) at (12.100002cm,-0.000000cm){};
\node[enqueue, minimum width=0.25cm](tsk_0_15) at (16.400002cm,-0.000000cm){};
\node[enqueue, minimum width=0.25cm](tsk_0_16) at (16.700001cm,-0.000000cm){};
\node[enqueue, minimum width=0.25cm](tsk_0_17) at (17.000000cm,-0.000000cm){};
\node[exchange, minimum width=0.75cm](tsk_0_18) at (17.599998cm,-0.000000cm){};
\node[walk, minimum width=1.65cm](tsk_0_19) at (18.399998cm,-0.000000cm){};
\node[cell, minimum width=0.75cm](tsk_0_20) at (20.099998cm,-0.000000cm){};
\node[enqueue, minimum width=0.25cm](tsk_0_21) at (20.899998cm,-0.000000cm){};
\node[enqueue, minimum width=0.25cm](tsk_0_22) at (21.199997cm,-0.000000cm){};
\node[enqueue, minimum width=0.25cm](tsk_0_23) at (21.499996cm,-0.000000cm){};
\node[enqueue, minimum width=0.25cm](tsk_0_24) at (21.799995cm,-0.000000cm){};
\node[enqueue, minimum width=0.25cm](tsk_0_25) at (22.099995cm,-0.000000cm){};

\node[cell, minimum width=0.75cm](tsk_1_0) at (0.000000cm,-0.700000cm){};
\node[cell, minimum width=0.75cm](tsk_1_1) at (0.800000cm,-0.700000cm){};
\node[cell, minimum width=0.75cm](tsk_1_2) at (1.600000cm,-0.700000cm){};
\node[cell, minimum width=0.75cm](tsk_1_3) at (2.400000cm,-0.700000cm){};
\node[cell, minimum width=0.75cm](tsk_1_4) at (3.200000cm,-0.700000cm){};
\node[enqueue, minimum width=0.25cm](tsk_1_5) at (4.000000cm,-0.700000cm){};
\node[enqueue, minimum width=0.25cm](tsk_1_6) at (4.300000cm,-0.700000cm){};
\node[enqueue, minimum width=0.25cm](tsk_1_7) at (4.600000cm,-0.700000cm){};
\node[cell, minimum width=0.75cm](tsk_1_8) at (5.050001cm,-0.700000cm){};
\node[cell, minimum width=0.75cm](tsk_1_9) at (5.850001cm,-0.700000cm){};
\node[cell, minimum width=0.75cm](tsk_1_10) at (6.650001cm,-0.700000cm){};
\node[cell, minimum width=0.75cm](tsk_1_11) at (7.450001cm,-0.700000cm){};
\node[cell, minimum width=0.75cm](tsk_1_12) at (8.250001cm,-0.700000cm){};
\node[enqueue, minimum width=0.25cm](tsk_1_13) at (9.050001cm,-0.700000cm){};
\node[enqueue, minimum width=0.25cm](tsk_1_14) at (9.350001cm,-0.700000cm){};
\node[enqueue, minimum width=0.25cm](tsk_1_15) at (9.650002cm,-0.700000cm){};
\node[cell, minimum width=0.75cm](tsk_1_16) at (10.000002cm,-0.700000cm){};
\node[cell, minimum width=0.75cm](tsk_1_17) at (10.800002cm,-0.700000cm){};
\node[cell, minimum width=0.75cm](tsk_1_18) at (11.600002cm,-0.700000cm){};
\node[cell, minimum width=0.75cm](tsk_1_19) at (12.400002cm,-0.700000cm){};
\node[cell, minimum width=0.75cm](tsk_1_20) at (13.200003cm,-0.700000cm){};
\node[enqueue, minimum width=0.25cm](tsk_1_21) at (16.350002cm,-0.700000cm){};
\node[enqueue, minimum width=0.25cm](tsk_1_22) at (16.650002cm,-0.700000cm){};
\node[enqueue, minimum width=0.25cm](tsk_1_23) at (16.950001cm,-0.700000cm){};
\node[enqueue, minimum width=0.25cm](tsk_1_24) at (17.250000cm,-0.700000cm){};
\node[cell, minimum width=0.75cm](tsk_1_25) at (17.599998cm,-0.700000cm){};
\node[cell, minimum width=0.75cm](tsk_1_26) at (18.399998cm,-0.700000cm){};
\node[cell, minimum width=0.75cm](tsk_1_27) at (19.199997cm,-0.700000cm){};
\node[cell, minimum width=0.75cm](tsk_1_28) at (19.999996cm,-0.700000cm){};
\node[cell, minimum width=0.75cm](tsk_1_29) at (20.799995cm,-0.700000cm){};
\node[enqueue, minimum width=0.25cm](tsk_1_30) at (21.599995cm,-0.700000cm){};
\node[enqueue, minimum width=0.25cm](tsk_1_31) at (21.899994cm,-0.700000cm){};
\node[enqueue, minimum width=0.25cm](tsk_1_32) at (22.199993cm,-0.700000cm){};

\node[cell, minimum width=0.75cm](tsk_2_0) at (0.000000cm,-1.400000cm){};
\node[cell, minimum width=0.75cm](tsk_2_1) at (0.800000cm,-1.400000cm){};
\node[cell, minimum width=0.75cm](tsk_2_2) at (1.600000cm,-1.400000cm){};
\node[cell, minimum width=0.75cm](tsk_2_3) at (2.400000cm,-1.400000cm){};
\node[cell, minimum width=0.75cm](tsk_2_4) at (3.200000cm,-1.400000cm){};
\node[enqueue, minimum width=0.25cm](tsk_2_5) at (4.000000cm,-1.400000cm){};
\node[enqueue, minimum width=0.25cm](tsk_2_6) at (4.300000cm,-1.400000cm){};
\node[enqueue, minimum width=0.25cm](tsk_2_7) at (4.600000cm,-1.400000cm){};
\node[cell, minimum width=0.75cm](tsk_2_8) at (5.050001cm,-1.400000cm){};
\node[cell, minimum width=0.75cm](tsk_2_9) at (5.850001cm,-1.400000cm){};
\node[cell, minimum width=0.75cm](tsk_2_10) at (6.650001cm,-1.400000cm){};
\node[cell, minimum width=0.75cm](tsk_2_11) at (7.450001cm,-1.400000cm){};
\node[cell, minimum width=0.75cm](tsk_2_12) at (8.250001cm,-1.400000cm){};
\node[enqueue, minimum width=0.25cm](tsk_2_13) at (9.050001cm,-1.400000cm){};
\node[enqueue, minimum width=0.25cm](tsk_2_14) at (9.350001cm,-1.400000cm){};
\node[enqueue, minimum width=0.25cm](tsk_2_15) at (9.650002cm,-1.400000cm){};
\node[cell, minimum width=0.75cm](tsk_2_16) at (10.000002cm,-1.400000cm){};
\node[cell, minimum width=0.75cm](tsk_2_17) at (10.800002cm,-1.400000cm){};
\node[cell, minimum width=0.75cm](tsk_2_18) at (11.600002cm,-1.400000cm){};
\node[cell, minimum width=0.75cm](tsk_2_19) at (12.400002cm,-1.400000cm){};
\node[cell, minimum width=0.75cm](tsk_2_20) at (13.200003cm,-1.400000cm){};
\node[enqueue, minimum width=0.25cm](tsk_2_21) at (16.350002cm,-1.400000cm){};
\node[enqueue, minimum width=0.25cm](tsk_2_22) at (16.650002cm,-1.400000cm){};
\node[enqueue, minimum width=0.25cm](tsk_2_23) at (16.950001cm,-1.400000cm){};
\node[enqueue, minimum width=0.25cm](tsk_2_24) at (17.250000cm,-1.400000cm){};
\node[cell, minimum width=0.75cm](tsk_2_25) at (17.599998cm,-1.400000cm){};
\node[cell, minimum width=0.75cm](tsk_2_26) at (18.399998cm,-1.400000cm){};
\node[cell, minimum width=0.75cm](tsk_2_27) at (19.199997cm,-1.400000cm){};
\node[cell, minimum width=0.75cm](tsk_2_28) at (19.999996cm,-1.400000cm){};
\node[cell, minimum width=0.75cm](tsk_2_29) at (20.799995cm,-1.400000cm){};
\node[enqueue, minimum width=0.25cm](tsk_2_30) at (21.599995cm,-1.400000cm){};
\node[enqueue, minimum width=0.25cm](tsk_2_31) at (21.899994cm,-1.400000cm){};

\node[cell, minimum width=0.75cm](tsk_3_0) at (0.000000cm,-2.100000cm){};
\node[cell, minimum width=0.75cm](tsk_3_1) at (0.800000cm,-2.100000cm){};
\node[cell, minimum width=0.75cm](tsk_3_2) at (1.600000cm,-2.100000cm){};
\node[cell, minimum width=0.75cm](tsk_3_3) at (2.400000cm,-2.100000cm){};
\node[enqueue, minimum width=0.25cm](tsk_3_4) at (3.200000cm,-2.100000cm){};
\node[enqueue, minimum width=0.25cm](tsk_3_5) at (3.500000cm,-2.100000cm){};
\node[enqueue, minimum width=0.25cm](tsk_3_6) at (3.800000cm,-2.100000cm){};
\node[enqueue, minimum width=0.25cm](tsk_3_7) at (4.100000cm,-2.100000cm){};
\node[enqueue, minimum width=0.25cm](tsk_3_8) at (4.400000cm,-2.100000cm){};
\node[enqueue, minimum width=0.25cm](tsk_3_9) at (4.700000cm,-2.100000cm){};
\node[cell, minimum width=0.75cm](tsk_3_10) at (5.050001cm,-2.100000cm){};
\node[cell, minimum width=0.75cm](tsk_3_11) at (5.850001cm,-2.100000cm){};
\node[cell, minimum width=0.75cm](tsk_3_12) at (6.650001cm,-2.100000cm){};
\node[cell, minimum width=0.75cm](tsk_3_13) at (7.450001cm,-2.100000cm){};
\node[enqueue, minimum width=0.25cm](tsk_3_14) at (8.250001cm,-2.100000cm){};
\node[enqueue, minimum width=0.25cm](tsk_3_15) at (8.550001cm,-2.100000cm){};
\node[enqueue, minimum width=0.25cm](tsk_3_16) at (8.850001cm,-2.100000cm){};
\node[enqueue, minimum width=0.25cm](tsk_3_17) at (9.150002cm,-2.100000cm){};
\node[enqueue, minimum width=0.25cm](tsk_3_18) at (9.450002cm,-2.100000cm){};
\node[cell, minimum width=0.75cm](tsk_3_19) at (10.000002cm,-2.100000cm){};
\node[cell, minimum width=0.75cm](tsk_3_20) at (10.800002cm,-2.100000cm){};
\node[cell, minimum width=0.75cm](tsk_3_21) at (11.600002cm,-2.100000cm){};
\node[cell, minimum width=0.75cm](tsk_3_22) at (12.400002cm,-2.100000cm){};
\node[cell, minimum width=0.75cm](tsk_3_23) at (13.200003cm,-2.100000cm){};
\node[enqueue, minimum width=0.25cm](tsk_3_24) at (16.350002cm,-2.100000cm){};
\node[enqueue, minimum width=0.25cm](tsk_3_25) at (16.650002cm,-2.100000cm){};
\node[enqueue, minimum width=0.25cm](tsk_3_26) at (16.950001cm,-2.100000cm){};
\node[enqueue, minimum width=0.25cm](tsk_3_27) at (17.250000cm,-2.100000cm){};
\node[cell, minimum width=0.75cm](tsk_3_28) at (17.599998cm,-2.100000cm){};
\node[cell, minimum width=0.75cm](tsk_3_29) at (18.399998cm,-2.100000cm){};
\node[cell, minimum width=0.75cm](tsk_3_30) at (19.199997cm,-2.100000cm){};
\node[cell, minimum width=0.75cm](tsk_3_31) at (19.999996cm,-2.100000cm){};
\node[enqueue, minimum width=0.25cm](tsk_3_32) at (20.799995cm,-2.100000cm){};
\node[enqueue, minimum width=0.25cm](tsk_3_33) at (21.099995cm,-2.100000cm){};
\node[enqueue, minimum width=0.25cm](tsk_3_34) at (21.399994cm,-2.100000cm){};
\node[enqueue, minimum width=0.25cm](tsk_3_35) at (21.699993cm,-2.100000cm){};
\node[enqueue, minimum width=0.25cm](tsk_3_36) at (21.999992cm,-2.100000cm){};

\draw[thick](0.000000cm,0.200000cm) -- (0.000000cm,-2.800000cm);
\node[anchor=south] at (0.000000cm,0.200000cm) {$t_{0}$};
\draw[thick](5.015000cm,0.200000cm) -- (5.015000cm,-2.800000cm);
\node[anchor=south] at (5.015000cm,0.200000cm) {$t_{1}$};
\draw[thick](9.965002cm,0.200000cm) -- (9.965002cm,-2.800000cm);
\node[anchor=south] at (9.965002cm,0.200000cm) {$t_{2}$};
\draw[thick](17.564999cm,0.200000cm) -- (17.564999cm,-2.800000cm);
\node[anchor=south] at (17.564999cm,0.200000cm) {$t_{3}$};
\draw[thick](22.514992cm,0.200000cm) -- (22.514992cm,-2.800000cm);
\node[anchor=south] at (22.514992cm,0.200000cm) {$t_{4}$};
     \end{tikzpicture}
    \caption{
    A timeline for task execution for 4 threads with 15 cells.
    Tasks from left to right on the first (top) thread:
    spike global communication; enqueue spike events; update cell state;
    merge per-cell event queues.
    The spike communication and enqueue tasks are serialized
    and the event merge tasks can only be executed
    when the enqueue is complete. Compute and communication can
    fully overlap only if the combined time for spike communication and enqueue
    does not exceed the time taken for state updates.
    Thread starvation is evident in the third epoch.
    }
    \label{fig:threads}
\end{figure*}
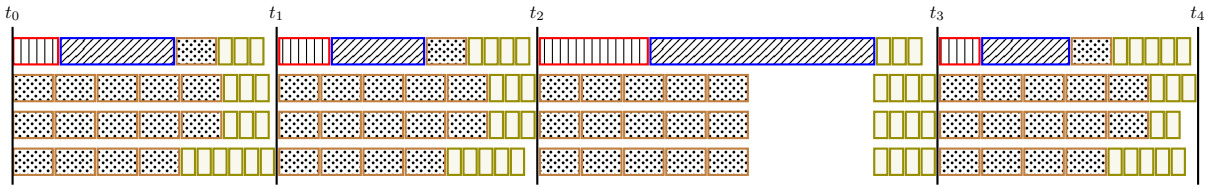

\arbor is a multi-threaded and distributed simulator; it uses a task-based
threading model (see \sect{sec:impl-taskpool} below) for the scheduling of cell
state integration tasks and for the communication of spike information across MPI
ranks.

\begin{figure}[h!]
    \includegraphics[width=0.47\textwidth]{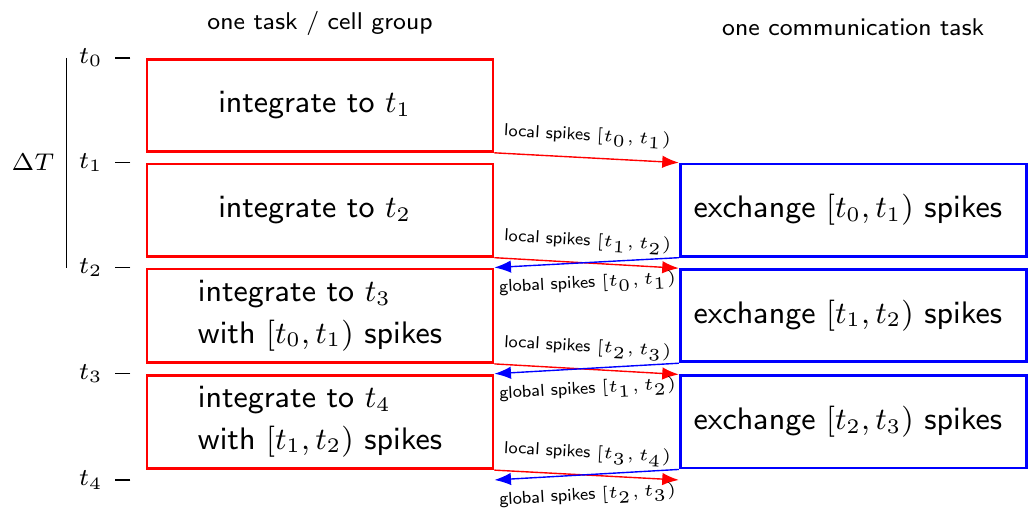}
    \caption{Overlapped compute and communication.}
    \label{fig:spike-overlap}
\end{figure}

When the simulation object is initialized, \arbor instantiates the cell group
data from the recipe in parallel, with each rank running one task per local
cell group as described in \sect{sec:design-groups}. The actual execution of
the simulation itself is broken into \emph{epochs}, each comprising a
simulation time interval of duration $\Delta T/2$ where $\Delta T$ is the
minimum propagation delay (see \sect{sec:model}). This propagation time allows
for the overlapping of state integration and spike exchange: the integration of
states in epoch $i$ requires the spikes generated in epoch $i-2$, and exchanged
in epoch $i-1$ (see \fig{fig:threads} and \fig{fig:spike-overlap}).
In each epoch, one task is run
for each cell group state integration, and one for the spike exchange.

Spike exchange consists of three steps: first, all spikes generated
in the previous epoch are sorted and then collected in each MPI rank
via an \textsc{allgather} and \textsc{allgatherv} operation; each spike
from the global list is checked against a table of local connections,
generating corresponding events on per-cell event queues; finally,
these event queues are merged with the pending events for each
cell in parallel. 

The local connection table is partitioned by the rank of the source cell, and
then sorted within each partition by destination cell. The global spike list is
also naturally partitioned by source rank by virtue of the \text{allgatherv}
operation; the testing of spikes against local connections is then
$O(S\cdot\log(NC/R))$ for homogeneous networks, where $S$ is the number of
spikes, $R$ the number of ranks, $N$ the number of cells per rank, and $C$ the
number of connections per cell.

\subsection{Task pool}
\label{sec:impl-taskpool}

We implemented a simple tasking system with dedicated task queues per thread,
task stealing and exception handling. Tasks are submitted to a task pool and
pushed to the thread queues in a round-robin fashion. Threads execute tasks
from their queues; if their queues are empty, they execute tasks from another
thread's non-empty, available queue.

Tasks are submitted in \emph{task groups}, within which they are processed
asynchronously, but joined synchronously. Exceptions generated by a task
within a task group are propagated to the caller when the task group
is joined.

The entire implementation is compact --- only 360 lines of code --- while its
performance is comparable to sophisticated threading libraries for
representative task sizes.

\subsection{Architecture-specific optimization}
\label{sec:impl-optimization}

\arbor employs specific optimizations for GPU and CPU implementations
in order to take advantage of the merits of each architecture.

As mentioned in \sect{sec:design-groups}, efficient utilization of
a GPU requires very wide parallelism. A GPU cell group for multi-compartment
cells will typically encapsulate all the cells on the rank, with
integration and event handling performed in-step.

Mechanism state update is highly parallel, with one CUDA thread per instance;
however the collection of the corresponding membrane currents in each
control volume from potentially many synapses and other sources must be
synchronized. \arbor uses a CUDA-intrinsics based key reduction algorithm
to limit the use of atomic operations in this summation.

The Thomas algorithm, as used in the implicit Euler step for the membrane
voltage update, is inherently serial. The GPU implementation uses one thread
per cell, but the matrix layouts for the cell group are interleaved
in order to maximize utilization of the GPU memory bandwidth.

\arbor uses explicit vectorization for CPU targets, with support for AVX, AVX2,
and AVX512. This vectorization support is provided by a dedicated vectorization
library based on the recent C++ standards proposal P0214R6~\cite{KretzProposal}
that separates the interface from the architecture-specific vectorization
intrinsics. The library additionally supports gather/scatter operations subject
to supplied constraints on the indirect indices and architecture-specific
vectorized implementations of a number of transcendental functions.
Support for additional vector instruction sets can be added in a modular way.

The \texttt{modcc} transpiler will translate \nmodl descriptions of ion channel and
synapse dynamics to vectorized kernels for the mechanism state update and
current contribution operations, which dominate time to solution on CPU.
The requisite data for these operations is not always sequential in
memory; synapses, in particular, may have a many-to-one relationship with
their corresponding control volumes in the cell discretizations.
Accessing this data via vector gather/scatter operations, however, is much slower
than, for example, a direct vector load or store operation~\cite{AgnerFog}.

\arbor performs a pre-processing step to factor the mechanism data accesses into
four indirection categories: sequential; constant, that is, accessing the one
location; independent non-sequential; and free. These constitute a constraint
on the indirect indices for the gather or scatter operations, which are then
performed in an architecture-optimized manner by the vectorization library.

\begin{figure}[t!]
    \begin{center}
        \begin{tikzpicture}
    [scale=0.9, every node/.style={scale=0.9}]
    \begin{axis}[
        ylabel=Speedup,
        ybar,
        width=0.5\textwidth,
        height=0.25\textwidth,
        xmin=50, xmax=450,
        ymin=1, ymax=4.5,
        ytick={1.0, 1.5, 2.0, 2.5, 3.0, 3.5, 4.0},
        xtick={100,200,300,400},
        xticklabels={Kaby Lake i7, Xeon Broadwell, Xeon Skylake, KNL},
        ylabel style={yshift=-10pt},
        yticklabel style={xshift=-2pt},
        legend style = {at={(0,1)}, anchor=north west},
        grid=major,
        tick label style={rotate=10},
        bar width=12pt]
              \addplot coordinates {
                (100,2.43) (200,1.54) (300,2.37) (400,4.05)};
              \addplot coordinates {
                (100,2.28) (200,1.56) (300,2.26) (400,3.40)};

       \legend{core, socket}
   \end{axis}
\end{tikzpicture}

         \caption{
            Single core and single socket speedup for explicit vectorization on four Intel architectures.
        }
        \label{fig:vector}
    \end{center}
\end{figure}
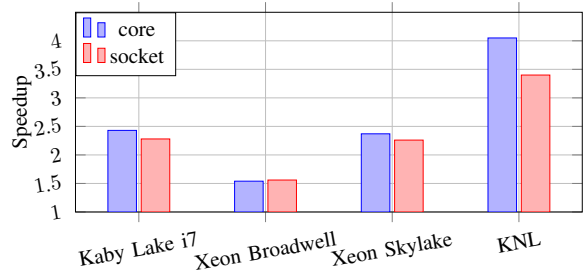

The use of the dedicated vectorization library allows for a significant
improvement in time to solution when compared with compiler auto-vectorized
code. We run a benchmark consisting of cells with 300 compartments with
Hodgkin-Huxley mechanisms and 5,000 randomly connected exponential synapses.
\fig{fig:vector} shows the per-core and per-socket
speedup for four Intel CPUs: a laptop Kaby Lake i7; a Broadwell Xeon; a Skylake Xeon; and KNL
(see \tbl{tbl:vector}). The use of data-pattern optimized loads and
stores contributes significantly to the observed gains; the comparatively
low improvement observed for the Xeon Broadwell is due to the poor
performance of vectorized division on this architecture, with half the
throughput of Kaby~Lake and Skylake-X~\cite{AgnerFog}.

\begin{table}[htp!]
    \begin{center}
        \scriptsize
        \caption{Vectorization benchmark CPU characteristics.}
        \begin{tabular}{|rrrr|}
            \hline\TS
            CPU          & cores & threads & ISA     \\
            \hline\TS
            Kaby Lake i7  & 2     & 4       & AVX-2   \\
            Broadwell    & 18    & 36      & AVX-2   \\
            Skylake-X    & 18    & 36      & AVX-512 \\
            KNL          & 64    & 256     & AVX-512 \\
            \hline
        \end{tabular}
        \label{tbl:vector}
    \end{center}
\end{table}

\section{Benchmarks}
The performance of a neural network simulation application can be measured in three ways
from a HPC user's perspective: the total time to solution; the allocation resource
usage (node or cpu-core hours to solution); and the maximum model size that can be simulated.

These measures are determined by single node performance, and the strong and weak scaling behavior of the application. \emph{Strong scaling} measures the performance gain obtained for a fixed problem size as the number of nodes is increased, and \emph{weak scaling} measures the time to solution as a problem size is increased in proportion to the number of nodes.

For \arbor to meet users' performance expectations, it has to fit and efficiently execute a large enough model on one node, and then weak scale well enough to simulate the entire problem.

We want to emphasize that the aim of the benchmark results presented here is not to say that one architecture is better than the other.
Instead our intention is to illustrate that \arbor can be used effectively on any HPC system available to a scientist.

\subsection{Test configuration}
\label{sec:systems}

\subsubsection{Test systems}

Three Cray clusters at CSCS were used for the benchmarks, detailed in
\tbl{tbl:nodes}. All benchmarks benefited from hyperthreading,
which was enabled in each configuration.

\begin{table}[htp!]
    \begin{center}
        \scriptsize
        \caption{ The resources on one node of each of the three test systems.  }
        \label{tbl:nodes}
        \begin{tabular}{|c|lll|}
            \cline{2-4}
\multicolumn{1}{c|}{}\TS &   \daintmc    &   \daintgpu   & \tave  \\
            \hline\TS
CPU                     &   Broadwell   &   Haswell     & KNL   \\
memory                  &   64 GB       &   32 GB       & 96 GB \\
CPU sockets             &   2           &   1           & 1      \\
cores/socket            &   18          &   12          & 64     \\
threads/core            &   2           &   2           & 4      \\
vectorization           &   AVX2        &   AVX2        & AVX512 \\
accelerator             &   --          &   P100 GPU    & --     \\
interconnect            &   Aries       &   Aries       & Aries  \\
MPI ranks               &   2           &   1           & 4      \\
threads/rank            &   36          &   24          & 64    \\
configuration           &   --          &   CUDA 9.2    & cache,quadrant \\
compiler                &   GCC 7.2.0   &   GCC 6.2.0   & GCC 7.2.0      \\
            \hline
        \end{tabular}
    \end{center}
\end{table}

\subsubsection{Software}

\arbor version 0.1;
\nrn version 7.6.2, with Python 3.
\arbor and \nrn compiled from source with Cray MPI.

\subsubsection{Models}
\label{sec:bench_models}

Two models are used to benchmark \arbor and \nrn.

\textbf{Ring model}:
Randomly generated cell morphologies with on average 130
compartments and 10,000 exponential synapses per cell, where only one of the
synapses is connected to a spike detector on the preceding cell to form a ring
network. The soma has Hodgkin-Huxley mechanics, and dendrites have passive
conductance. The model is useful for evaluating the computational overheads of
cell updates without any spike communication or network overheads.

\textbf{10k connectivity model}:
Using the same cells as for the ring model,
with a 10,000-way randomly connected network with no self-connections. The
minimum delay is either 10ms or 20ms, and all synapses are excitatory. All
cells spike synchronously with frequency 50Hz, which is a pathological
case for spike communication, making this model useful for testing the
scalability of spike communication.

Validation of model outputs is not described in this benchmark paper due to space constraints.
Briefly, the results were validated by first comparing voltage traces of individual cells.
Then spike trains of the respective models were compared, which is trivial for the ring model,
and required ensuring that the synchronous spiking occurred with the same frequency and phase
in the 10k connectivity model. Validation of individual cell models and network spikes is
on-going work, that will be made available online in an open source validation and benchmarking
suite.

\subsection{Single node performance}
\label{sec:single}

\tbl{tbl:single} shows the time to solution and energy overheads \cite{fourestey-2018} of the ring model with 32--16,384 cells on a single node of \daintmc, \daintgpu and \tave.

The nodes on the test sytems offer range of computation resources, from 76 parallel threads on \daintmc, to 256 threads on \tave, and thousands of GPU cores on \daintgpu.
Time to solution decreases on each system as the number of cells is reduced, however these gains are marginal for model sizes below a system-dependent threshold,
and this threshold is higher for systems with more on-node parallelism.
Efficient use of \daintmc requires a minimum of 64 cells; for \tave, 512 cells; and for \daintgpu, 1024 cells.

\begin{table}[htp!]
    \begin{center}
        \scriptsize
        \caption{Single node results. \daintmc 2 MPI ranks with 36 threads, \daintgpu 1 rank with 24 threads, \tave cache mode with 4 MPI ranks with 64 threads (4 per core).}
        \begin{tabular}{|r|rrrr|rrr|}
            \cline{2-8}
             \multicolumn{1}{c|}{}\TS  & \multicolumn{4}{c|}{wall time (s)} & \multicolumn{3}{c|}{energy (kJ)}  \\
            \hline\TS
cells & mc & gpu & knl & nrn & mc & gpu & knl \\
            \hline\TS
   32  & \hl{ 0.35} &      2.06  &  1.13 &    1.73 & \hl{0.04} &     0.25  &  0.17 \\
   64  & \hl{ 0.39} &      2.10  &  1.29 &    2.61 & \hl{0.05} &     0.25  &  0.22 \\
  128  & \hl{ 0.75} &      2.44  &  1.71 &    8.27 & \hl{0.11} &     0.33  &  0.34 \\
  256  & \hl{ 1.42} &      2.97  &  2.28 &   32.92 & \hl{0.26} &     0.43  &  0.55 \\
  512  & \hl{ 2.66} &      4.19  &  3.36 &   67.33 & \hl{0.58} &     0.67  &  0.97 \\
 1024  & \hl{ 5.12} &      6.50  &  6.15 &  135.52 &     1.24  & \hl{1.14} &  1.81 \\
 2048  & \hl{10.04} &     11.11  & 12.27 &  272.87 &     2.53  & \hl{2.11} &  3.63 \\
 4096  & \hl{19.93} &     19.96  & 24.39 &  555.34 &     5.16  & \hl{3.96} &  7.24 \\
 8192  &     39.66  & \hl{37.24} & 48.65 & 1234.70 &    10.38  & \hl{7.72} & 14.45 \\
16384  &     79.22  & \hl{71.65} & 97.19 &    --   &    20.85  &\hl{15.11} & 28.99 \\
            \hline
        \end{tabular}
        \label{tbl:single}
    \end{center}
\end{table}

\subsection{Comparison with \nrn}
\label{sec:nrn}

\nrn~\cite{Carnevale2006} is the most widely used software for general simulation of networks of multi-compartment cells.
Like \arbor, it supports running large models in parallel using multi-threading and MPI~\cite{brette-2007}.
We use NEURON for comparison because of its ubiquity, because of its support for distributed execution, because it is under active development, and because, like Arbor, it is designed for models with user-defined cell types, synapses and ion channels.

\begin{figure}[ht]
    \centering
    \begin{tikzpicture}
    [scale=0.9, every node/.style={scale=0.9}]
    \begin{axis}[
        xmode=log,
        height=0.25\textwidth,
        width=0.5\textwidth,
        xmin=64,xmax=8192,
        ymin=0, ymax=35,
        xtick={64, 128, 256, 512, 1024, 2048, 4096, 8193, 16386},
        xticklabels={64, 128, 256, 512, 1k, 2k, 4k, 8k, 16k},
        ytick={0,5,10,15,20,25,30,35},
        ylabel=speedup,
        xlabel=cells,
        xticklabel style={yshift=-2pt},
        yticklabel style={xshift=-2pt},
        legend style = {at={(1,0.2)}, anchor=south east},
        line width=1pt,
        every axis y label/.style=
            {at={(ticklabel cs:0.5)},rotate=90,anchor=near ticklabel},
        grid=major]

        \addplot[color=blue, mark=*, mark size=2, mark options={fill=white}]
            table[x=cells,y expr=\thisrow{nrnmc_wall}/\thisrow{arbmc_wall}] {./data/nrn_arb_single.tbl};
        \addplot[color=red, mark=triangle, mark size=2, mark options={fill=white}]
            table[x=cells,y expr=\thisrow{nrnmc_wall}/\thisrow{arbgpu_wall}] {./data/nrn_arb_single.tbl};
        \addplot[color=black!50!white]
            table[x=cells,y expr=\thisrow{nrnmc_wall}/\thisrow{nrnmc_wall}] {./data/nrn_arb_single.tbl};

       \legend{
                {\scriptsize \arbor{}-mc},
                {\scriptsize \arbor{}-gpu},
              };
    \end{axis}
\end{tikzpicture}

     \caption{The single node speedup of \arbor running on \daintmc and \daintgpu relative to \nrn on \daintmc.}
    \label{fig:arb-v-nrn-speedup}
\end{figure}
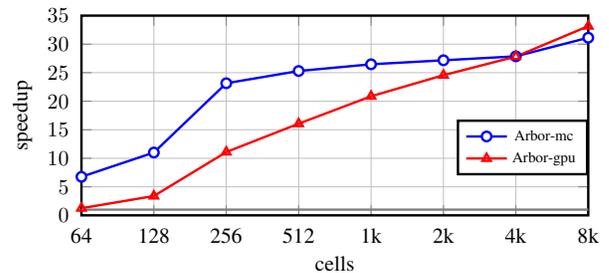

The wall time for simulating the ring benchmark with \nrn on \daintmc is also tabulated in \tbl{tbl:single}.
\arbor is faster than \nrn for all model sizes, with the speedup increasing with model size (see \fig{fig:arb-v-nrn-speedup}).
For fewer than 128 cells \arbor is $5$--$10\times$ faster, and for more than 256 cells it is over $20\times$ faster.
\arbor is also significantly more memory efficient: the 16k cell model required 4.4 GB of memory, whereas \nrn was unable to run
this model in the 64 GB memory available on \daintmc.

These gains are primarily due to more efficient memory bandwidth and cache use in \arbor: \arbor uses a
structure-of-array (SoA) memory layout, as opposed to the array-of-structure layout of \nrn, and for larger
models, \arbor is able to keep cell state in L2 and L3 cache where \nrn becomes DRAM-bandwidth bound.

\subsection{Strong Scaling}
\label{sec:strong}

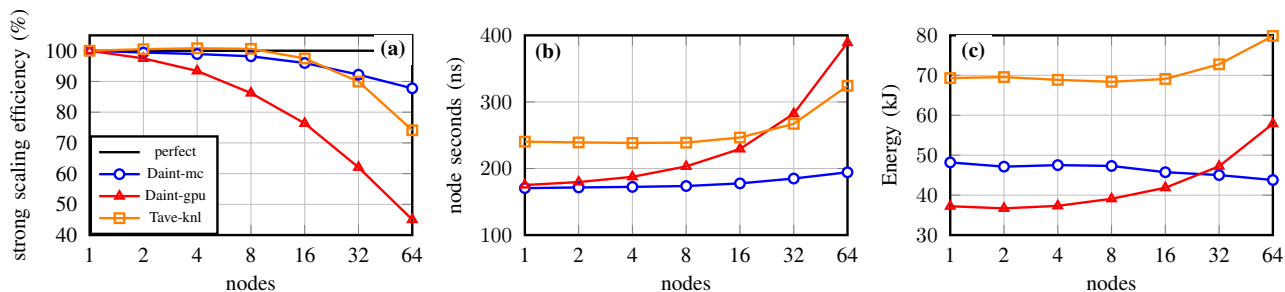
\begin{figure*}[htp!]
    \centering
    \begin{tikzpicture}
    [scale=0.9, every node/.style={scale=0.9}]
    \begin{axis}[
        xmode=log,
        height=0.25\textwidth,
        width=0.35\textwidth,
        xmin=1, xmax=64,
        xtick={1, 2, 4, 8, 16, 32, 64},
        xticklabels={1, 2, 4, 8, 16, 32, 64},
        ymin=0.4, ymax=1.05,
        ytick={0, 0.1, 0.2, 0.3, 0.4, 0.5, 0.6, 0.7, 0.8, 0.9, 1},
        yticklabels={0, 10, 20, 30, 40, 50, 60, 70, 80, 90, 100},
        ylabel=strong scaling efficiency (\%),
        xlabel=nodes,
        xticklabel style={yshift=-2pt},
        yticklabel style={xshift=-2pt},
        legend style = {at={(0,0)}, anchor=south west},
        line width=1pt,
        every axis y label/.style=
            {at={(ticklabel cs:0.5)},rotate=90,anchor=near ticklabel},
        grid=major]

        \addplot[color=black]
            table[x=nodes,y=p_eff] {data/strong_multi.tbl};
        \addplot[color=blue, mark=*, mark size=2, mark options={fill=white}]
            table[x=nodes,y=mc_eff] {data/strong_multi.tbl};
        \addplot[color=red, mark=triangle, mark size=2, mark options={fill=white}]
            table[x=nodes,y=gpu_eff] {data/strong_multi.tbl};
        \addplot[color=orange, mark=square, mark size=2, mark options={fill=white}]
            table[x=nodes,y=knl_eff] {data/strong_multi.tbl};

        \node[above, fill=white, align=center, inner sep=1mm]
              (a) at (axis cs:50,0.95){\textbf{(a)}};

       \legend{ {\scriptsize perfect},
                {\scriptsize \daintmc},
                {\scriptsize \daintgpu},
                {\scriptsize \tave},
              };
    \end{axis}
    \end{tikzpicture}
     \begin{tikzpicture}
    [scale=0.9, every node/.style={scale=0.9}]
    \begin{axis}[
        xmode=log,
        height=0.25\textwidth,
        width=0.35\textwidth,
        xmin=1, xmax=64,
        xtick={1, 2, 4, 8, 16, 32, 64},
        xticklabels={1, 2, 4, 8, 16, 32, 64},
        ymin=100, ymax=400,
                ylabel=node seconds (ns),
        xlabel=nodes,
        xticklabel style={yshift=-2pt},
        yticklabel style={xshift=-2pt},
        line width=1pt,
        every axis y label/.style=
            {at={(ticklabel cs:0.5)},rotate=90,anchor=near ticklabel},
        grid=major]

        \addplot[color=blue, mark=*, mark size=2, mark options={fill=white}]
            table[x=nodes,y expr=\thisrow{mc_time}*\thisrow{nodes}] {data/strong_multi.tbl};
        \addplot[color=red, mark=triangle, mark size=2, mark options={fill=white}]
            table[x=nodes,y expr=\thisrow{gpu_time}*\thisrow{nodes}] {data/strong_multi.tbl};
        \addplot[color=orange, mark=square, mark size=2, mark options={fill=white}]
            table[x=nodes,y expr=\thisrow{knl_time}*\thisrow{nodes}] {data/strong_multi.tbl};

        \node[above, fill=white, align=center, inner sep=1mm]
              (b) at (axis cs:1.35,350){\textbf{(b)}};
    \end{axis}
\end{tikzpicture}
     \begin{tikzpicture}
    [scale=0.9, every node/.style={scale=0.9}]
    \begin{axis}[
        xmode=log,
        height=0.25\textwidth,
        width=0.35\textwidth,
        xmin=1, xmax=64,
        xtick={1, 2, 4, 8, 16, 32, 64},
        xticklabels={1, 2, 4, 8, 16, 32, 64},
        ymin=30, ymax=80,
        ytick={30,40,50,60,70,80},
        ylabel=Energy (kJ),
        xlabel=nodes,
        xticklabel style={yshift=-2pt},
        yticklabel style={xshift=-2pt},
        line width=1pt,
        every axis y label/.style=
            {at={(ticklabel cs:0.5)},rotate=90,anchor=near ticklabel},
        grid=major]

        \addplot[color=blue, mark=*, mark size=2, mark options={fill=white}]
            table[x=nodes,y=mc_energy] {data/strong_multi.tbl};
        \addplot[color=red, mark=triangle, mark size=2, mark options={fill=white}]
            table[x=nodes,y=gpu_energy] {data/strong_multi.tbl};
        \addplot[color=orange, mark=square, mark size=2, mark options={fill=white}]
            table[x=nodes,y=knl_energy] {data/strong_multi.tbl};

        \node[above, fill=white, align=center, inner sep=1mm]
              (c) at (axis cs:1.35,72){\textbf{(c)}};
    \end{axis}
\end{tikzpicture}
     \caption{
        Strong scaling from 1 to 64 nodes of a 100~ms simulation with 16,384 cells and 10,000 randomly connected synapses.
        Efficiency decreases as the number of nodes increases, but only the multi-core system scales with 90\% efficiency to 64 nodes (256 cells per node) in \textbf{(a)}.
        The resources consumed, in terms of node-seconds (simulation time $\times$ nodes) and total energy to solution are shown in \textbf{(b)} and \textbf{(c)} respectively.}
    \label{fig:strong_multi}
\end{figure*}

Strong scaling is closely related to single node performance.
The number of cells per node decreases as the number of nodes increases, and it follows from the single-node scaling results in~\sect{sec:single} that for each architecture there is a maximum number of nodes beyond which there are too few cells per node to utilize on-node resources.
It makes little sense to scale a model far past this point, because though time to solution decreases, the total CPU or node hours from an allocation increases.

\begin{table}[htp!]
    \begin{center}
        \scriptsize
        \caption{Strong scaling of 16,384k cell model up to 64 nodes of \daintmc (mc), \daintgpu (gpu) and \tave (knl).}
        \begin{tabular}{|r|r|rrr|rrr|}
             \cline{3-8}
             \multicolumn{2}{c|}{}\TS  & \multicolumn{3}{c|}{wall time (s)} & \multicolumn{3}{c|}{energy (kJ)}  \\
            \hline\TS
nodes & cell/node & mc & gpu & knl & mc & gpu & knl  \\
            \hline\TS
 1 &  16384 & 170.52  &  175.16 & 240.33 & 48.2  &  37.2 & 69.3\\
 2 &   8192 &  85.77  &   89.82 & 119.57 & 47.1  &  36.7 & 69.6\\
 4 &   4096 &  43.09  &   46.87 &  59.60 & 47.5  &  37.3 & 68.9\\
 8 &   2048 &  21.71  &   25.39 &  29.86 & 47.3  &  39.1 & 68.4\\
16 &   1024 &  11.10  &   14.34 &  15.41 & 45.7  &  41.9 & 69.1\\
32 &    512 &   5.78  &    8.82 &   8.35 & 45.0  &  47.2 & 72.8\\
64 &    256 &   3.04  &    6.08 &   5.07 & 43.8  &  57.9 & 79.9\\
            \hline
        \end{tabular}
        \label{tbl:strong_multi}
    \end{center}
\end{table}

\tbl{tbl:strong_multi} shows the time to solution and energy consumption for a 16,384 cell model with 10k connectivity run using 1 to 64 nodes.
The multicore and GPU nodes on \daintmc and \daintgpu respectively are equivalent (within 10\%) for 4000 or more cells per node (less than 4 nodes), and \daintmc is much more efficient for more nodes. A KNL node is uniformly slower than multicore, using 1.4$\times$ more time and energy.

Users of HPC systems are concerned with getting the most from their resource allocations.
The resource allocation overhead of running the same model each system, measured in node-seconds, is plotted in \fig{fig:strong_multi}(b). The energy to solution, which is more interesting for data centers, is shown in \fig{fig:strong_multi}(c).
These plots illustrate that resource utilization is efficient where strong scaling is efficient.
Hence both multicore and GPU systems are efficient for models with many cells on each node, and multicore systems can be used to more aggressively strong scale a model to minimize time to solution.

\subsection{Weak Scaling}
\label{sec:weak}
Weak scaling measures efficiency as the number of nodes is increased with a fixed amount of work per node.
Good weak scaling is a prerequisite for running large models efficiently on many nodes.

The weak scaling benchmarks presented here show that \arbor weak scales perfectly to hundreds of nodes on \daintmc and \daintgpu, and scales efficiently for very large models that might be run on the largest contemporary HPC systems. Previous studies have shown that \nrn also weak scales well to hundreds of nodes~\cite{Migliore2006}, hence for the sake of brevity \nrn weak scaling is not presented here.

\begin{figure}[htp!]
    \begin{center}
        \begin{tikzpicture}
    [scale=0.9, every node/.style={scale=0.9}]
    \begin{axis}[
                xmode=log,
        height=0.2\textwidth,
        width=0.45\textwidth,
        xmin=1,xmax=128,
        ymin=80, ymax=90,
                        xtick={1, 2, 4, 8 , 16, 32, 64, 128},
        xticklabels={1, 2, 4, 8 , 16, 32, 64, 128},
        ylabel=wall time (s),
        xlabel=nodes,
                xticklabel style={yshift=-2pt},
        yticklabel style={xshift=-2pt},
        legend style = {at={(1,0)}, anchor=south east},
        line width=1pt,
        every axis y label/.style=
            {at={(ticklabel cs:0.5)},rotate=90,anchor=near ticklabel},
        grid=major]

        \addplot[color=blue, mark=*, mark size=1.5, mark options={fill=white}]
            table[x=nodes,y=mc_wall] {./data/weak_real.tbl};
        \addplot[color=red, mark=*, mark size=1.5, mark options={fill=white}]
            table[x=nodes,y=gpu_wall] {./data/weak_real.tbl};

       \legend{ {\scriptsize \daintmc},
                {\scriptsize \daintgpu},
              };
    \end{axis}
\end{tikzpicture}
         \caption{
            Simulation time for the weak scaling tests with 8,192 cells per node, 1 to 128 nodes.
            Each cells is connected to 10,000 random cells with no self-connections.
        }
        \label{fig:weak_tts}
    \end{center}
\end{figure}
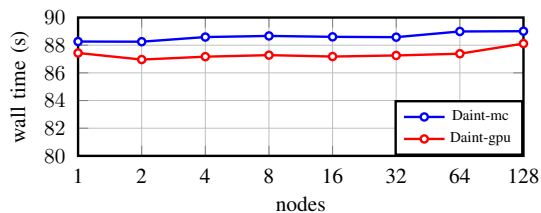

\fig{fig:weak_tts} shows the time solution for a model with 8,192 cells per node weak scaled from one to 1,048,576 cells on 128 nodes.
Both the GPU and CPU back ends have near perfect weak scaling for this range of nodes.
Piz Daint is a very busy system, and it isn't practical to extend the benchmark to thousands of nodes.

To explore scaling issues at larger scales, \arbor has a \emph{dry run} mode similar to the NEST simulator~\cite{Krunkel2017}.
Dry run mode runs a model on a single MPI rank, and mimics running on a large cluster by generating proxy spikes from cells on other ranks.
This approach allows us to investigate scaling on a larger numbers of nodes than otherwise practical or possible.

We use dry run mode with 36 threads on an 18 core socket of \daintmc to predict and model weak scaling when running at extreme scale.
The largest homogeneous Cray XC-40 system similar to \daintmc is the 9,688-node Cori at NERSC.
With this in mind 10,000 nodes was chosen as a reasonable maximum cluster size, and we test with 1,000 and 10,000 cells per node for a total model size of 10 million and 100 million cells respectively on 10,000 nodes.

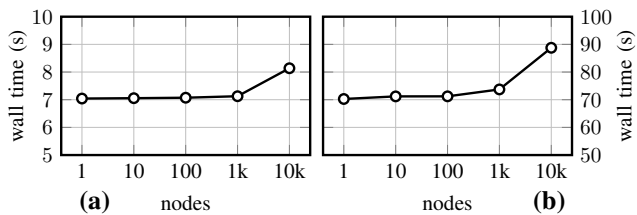
\begin{figure}[htp!]
    \begin{center}
            \begin{tikzpicture} [scale=0.9, every node/.style={scale=0.9}]
        \begin{axis}[
            height=0.2\textwidth,
            width=0.29\textwidth,
            xtick=      {1,  2,   3,    4,     5},
            xticklabels={1, 10, 100, {1k}, {10k}},
            ymin=5, ymax=10,
            ytick={0, 1, 2, 3, 4, 5, 6, 7, 8, 9, 10},
            legend style = {at={(0,0)}, anchor=south west},
            line width=1pt,
            ylabel=wall time (s),
            xlabel=nodes,
            every axis y label/.style=
                {at={(ticklabel cs:0.5)},rotate=90,anchor=near ticklabel,yshift=-5pt},
            grid=major]
            \addplot[color=black, mark=*, mark size=2, mark options={fill=white}]
                table[x=pos,y=c1k] {data/weak_wall.tbl};

        \end{axis}
    \end{tikzpicture}
    \begin{tikzpicture} [scale=0.9, every node/.style={scale=0.9}]
        \begin{axis}[
            height=0.2\textwidth,
            width=0.29\textwidth,
            xtick=      {1,  2,   3,    4,     5},
            xticklabels={1, 10, 100, {1k}, {10k}},
            ymin=50, ymax=100,
            ytick={0, 10, 20, 30, 40, 50, 60, 70, 80, 90, 100},
            yticklabel pos=right,
            legend style = {at={(0,0)}, anchor=south west},
            line width=1pt,
            ylabel=wall time (s),
            xlabel=nodes,
            every axis y label/.style=
                {at={(ticklabel cs:0.5)},rotate=90,anchor=near ticklabel,yshift=5pt},
            grid=major]
            \addplot[color=black, mark=*, mark size=2, mark options={fill=white}]
                table[x=pos,y=c10k] {data/weak_wall.tbl};
        \end{axis}
    \end{tikzpicture}
         \\ \vspace{-15pt}\textbf{(a)}\hspace{160pt}\textbf{(b)}\\
        \caption{
            Time to solution in seconds for the dry run weak scaling tests from 1 to 10,000 nodes for \textbf{(a)}
            1,000 cells per node and \textbf{(b)} 10,000 cells per node.
        }
        \label{fig:dry_tts}
    \end{center}
\end{figure}

The 100\,ms simulations have a 10\,ms min-delay, for 20 integration epochs (see \sect{sec:impl-workflow}), with cells firing at 87.5\,Hz.
\fig{fig:dry_tts} shows that the 1k and 10k models weak scale very well with 99\% and 95\% efficiency respectively at 1,000 nodes.
Weak scaling is still good at 10,000 nodes, 87\% and 79\% respectively, however it has clearly started to deteriorate.

\begin{table}[htp!]
    \begin{center}
        \scriptsize
        \caption{
            The accumulated time spent by all threads on each rank performing tasks for the dry-run benchmarks with 1k cells per node (top) and 10k cells per node (bottom).
            The costs serialized are highlighted in bold.
        }
        \begin{tabular}{|l|rrrrr|}
            \cline{2-6}
\multicolumn{1}{c|}{}  & \multicolumn{5}{c|}{nodes}\TS  \\
            \cline{1-6}\TS
region         &        1   &      10   &     100   &   1,000   &  10,000 \\
   \hline\TS
communication  & \hl{0.00}  &\hl{0.00}  &\hl{0.01}  &\hl{0.09}  &\hl{1.31}  \\
enqueue        & \hl{1.76}  &\hl{1.84}  &\hl{2.05}  &\hl{2.56}  &\hl{2.90}  \\
merge          &    5.49    &   6.15    &   6.25    &   6.17    &   6.20  \\
cell state     &   204.50   &  204.14   &  204.47   &  204.25   &  204.19 \\
idle           &    41.76   &   41.86   &   41.81   &   43.48   &   78.34 \\
\textsc{Total} &   253.51   &  253.98   &  254.59   &  256.54   &  292.93 \\

   \hline\TS
communication  & \hl{0.01}  &\hl{0.01}  &\hl{0.10}  &\hl{1.27}  &\hl{12.80} \\
enqueue        & \hl{23.63} &\hl{24.14} &\hl{26.31} &\hl{31.02} &\hl{36.80} \\
merge          &    62.65   &   63.33   &   63.88   &   62.87   &   62.94 \\
cell state     &  2070.69   & 2092.82   & 2086.59   & 2072.28   & 2089.95 \\
idle           &   372.45   &  382.93   &  387.07   &  486.44   &  993.02 \\
\textsc{Total} &  2529.43   & 2563.24   & 2563.96   & 2653.88   & 3195.50 \\
            \hline
        \end{tabular}
        \label{tbl:dry_weak}
    \end{center}
\end{table}

The overheads of the cell state and event merging tasks are fixed in \tbl{tbl:dry_weak},
which is reasonable given that the number of cells and post-synaptic spike events per cell are scale invariant when weak scaling this model.
On the other hand, the number of spikes that must be communicated and processed on each node to generate post-synaptic spike events increases proportionally to the number of nodes.
This is significant, because the spike communication and event enqueue tasks are serialized, and are not parallelized over multiple threads.

The optimized spike walking algorithm outlined in~\sect{sec:impl-workflow} scales well: it takes 1.5 times longer on 10,000 nodes than 1 node, despite walking 10,000 times more spikes.
However, the spike communication task scales linearly\footnote{The communication task in dry run generates a global spike vector filled with proxy spikes in
place of an MPI collective, so the timings are also a proxy for the actual MPI communication overheads.
We aim to improve this part of dry run mode by generating realistic timings with a performance model
for MPI collectives.},
so that the combined time for these serialized tasks doubles from 1 to 10,000 nodes for both models.

\fig{fig:threads} shows that when the combined time for communication and event enqueue tasks on one thread exceeds the time taken by other threads to update cell state, the other threads are blocked in idle state, which is 35 threads on \daintmc.
This model has highly synchronized spiking in 9 waves over 20 integration intervals, so the exchange and spike walk task time contributions are concentrated on some of the epochs.
The effect of this is clear, with the increased idle thread time and simulation time in \tbl{tbl:dry_weak} and \fig{fig:dry_tts} respectively.

Weak scaling of 80\% at extreme scale is considered satisfactory by the \arbor developers, given that the largest models being run in the \hbp are in the range 1 to 10 million cells.
When the need arises, the spike walk can be parallelized, however this will only be implemented when users require it to keep with \arbor's policy of avoiding premature optimization.

\section{Conclusion}
We have presented \arbor, a performance portable library for simulation of large networks of multi-compartment neurons on HPC systems.
The performance portability is by virtue of back-end specific optimizations for x86 multicore, Intel KNL, and NVIDIA GPUs.
These optimizations and low memory overheads make \arbor an order of magnitude faster than the most widely-used comparable simulation software.
The single-node performance can be scaled out to run very large models at extreme scale with efficient weak scaling.

\arbor is an open source project under active development.
Examples of new features that will be released soon include but are not limited to:
    a python wrapper for user-friendly model building and execution;
    accurate and efficient treatment of gap junctions;
    and a GPU solver for Hines matrices exposes more fine-grained parallelism.

\section*{Acknowledgment}
This research has received funding from the European Union’s Horizon 2020 Framework Programme for Research and Innovation under the Specific Grant Agreement No. 720270 (Human Brain Project SGA1), and Specific Grant Agreement No. 785907 (Human Brain Project SGA2).

\balance
\addtolength\bibitemsep{2pt plus 0.3ex}
\printbibliography
\vskip1pt

\end{document}